\documentclass[aps,12pt,superscriptaddress,amsfonts,amssymb,amsmath]{revtex4}
\pdfoutput=1

\usepackage{graphicx}
\usepackage{epsfig}
\usepackage{makeidx}
\usepackage{epstopdf}
\usepackage{natbib}
\usepackage{xcolor}
\usepackage{subcaption}
\usepackage{amsmath}

\begin{document}

\title{Are current discontinuities in molecular devices experimentally observable?}

\author{F.\ Minotti \footnote{Email address: minotti@df.uba.ar}}
\affiliation{Universidad de Buenos Aires, Facultad de Ciencias Exactas y Naturales, Departamento de F\'{\i}sica, Buenos Aires, Argentina}
\affiliation{CONICET-Universidad de Buenos Aires, Instituto de F\'{\i}sica del Plasma (INFIP), Buenos Aires, Argentina}

\author{G.\ Modanese \footnote{Email address: giovanni.modanese@unibz.it}}
\affiliation{Free University of Bozen-Bolzano \\ Faculty of Science and Technology \\ I-39100 Bolzano, Italy}

\linespread{0.9}

\begin{abstract}

An ongoing debate in the first-principles description of conduction in molecular devices concerns the correct definition of current in the presence of non-local potentials. If the physical current density ${\bf j}=(-ie\hbar/2m)(\Psi^* \nabla \Psi- \Psi \nabla \Psi^*)$ is not locally conserved but can be re-adjusted by a non-local term, which current should be regarded as real? We prove that the extended Maxwell equations by Aharonov-Bohm give the e.m.\ field generated by such currents without any ambiguity. For an oscillating dipole we show that the radiated electrical field has a longitudinal component proportional to $ \omega \hat{P}$, where $\hat{P}$ is the anomalous moment $\int \hat{I}(\mathbf{x})\mathbf{x} d^3x$ and $\hat{I}$ is the space-dependent part of the anomaly $I=\partial_t \rho+\nabla \cdot \mathbf{j}$. In the case of a stationary current in a molecular device, a failure of local current conservation causes a ``missing field'' effect that can be experimentally observable, especially if its entity depends on the total current.

\end{abstract}

\maketitle

\section{Introduction}

There are different opinions in the literature concerning the possible existence and physical  interpretation of quantum systems for which the electric charge is not locally conserved, i.e., the continuity condition $\partial_t \rho + \nabla \cdot \mathbf{j}=0$ is not valid everywhere.

Wang et al.\ \cite{li2008definition,zhang2011first} point out explicitly that in first-principles calculations of the current in molecular devices based on single-particle non-equilibrium Green’s function (NEGF) and density-functional theory
(DFT) the usual quantum-mechanical current ${\bf j}=(-ie\hbar/2m)(\Psi^* \nabla \Psi- \Psi \nabla \Psi^*)$ does not satisfy the continuity condition and must be complemented by a non-local term. The recourse to the NEGF-DFT formalism is necessary in order to take into account the contribution of the inner atomic shells. In principle a full quantum field theory of the system, if it can be formulated in a standard way including all the internal electrons, would yield a conserved \emph{bare} current; in practice, such a formulation does not exist for realistic systems \cite{lai2019charge}, and even if it could be achieved, it cannot be excluded that it needs to be renormalized, giving rise to quantum anomalies (effective breaking of symmetries and local conservation properties of the bare theory, like for the ABJ anomaly in condensed matter \cite{cheng1984gauge,parameswaran2014probing}).

An appealing feature of the approach by Wang et al.\ is that the additional current they propose is exactly the same as that originating from the extended Maxwell equations of Aharonov-Bohm without any consideration of the microscopic/quantum aspects. These equations represent essentially the only possible covariant extension of Maxwell's theory which is compatible with the established standard phenomenology of classical electromagnetism and QED and is also applicable to currents that are not locally conserved \cite{ohmura1956new, aharonov1963further,alicki1978generalised, cornille1990propagation, van2001generalisation, jimenez2011cosmological, hively2012toward, Modanese2017MPLB, modanese2017electromagnetic, arbab2017extended, hively2019classical,reed2020implications}.

In a series of papers starting in 2108, Jensen, Garner and collaborators have analysed in depth the concept of density current in quantum transport, applying it to specific molecules and obtaining results compatible with the approach by Wang et al.
Ref.\ \cite{cabra2018simulation} sets the general theoretical framework and 
notes that besides legitimate physical reasons for the non-conserving character of local currents, there are technical problems related to the choice of the basis in the first-principles calculations.
The discussion is illustrated by simulating elastic and
inelastic local currents in a benzenedithiol junction. Simulations
show that the local flux does not necessarily follow molecular bonds, with significant part of the flux going ``through space``.

In Ref.\ \cite{jensen2019current}, the current density is investigated in
saturated chains of alkanes, silanes and germanes.
The current density is defined in this context as
\begin{equation}
    \mathbf{j}(\mathbf{x})=\frac{e\hbar}{4m\pi} \sum_{i,j} \int_{-\infty}^{+\infty} dE \, G^<_{ij}(E) [\psi_i(\mathbf{x}) \nabla \psi_j(\mathbf{x})-\psi_j(\mathbf{x}) \nabla \psi_i(\mathbf{x})],
\end{equation}
where $G^<_{ij}(E)$ are the matrix elements of the lesser Green function in a standard non-orthogonal basis $\{ \psi_i \}$.
The authors show that an enlargement of the eigenbasis for the \emph{ab initio} calculations does not substantially improve the conservation of current density in this case (while it does lead to an improvement in other cases, notably for graphene ribbons \cite{walz2015local}). They then apply the recipe by Wang et al.\ computing the secondary currents via a Poisson equation.

In Refs.\ \cite{garner2019helical}
and \cite{garner2020three} linear carbon wires are considered. Detailed plots of the integrated local current density are given, as compared to the (constant) total current.  

There are also wave equations in quantum mechanics which do not derive from a microscopic theory but are proposed as effective models with several important applications, in which the current is not locally conserved \cite{lenzi2008solutions, lenzi2008fractional, latora1999superdiffusion, caspi2000enhanced, chamon1997nonlocal, balantekin1998green, laskin2002fractional, wei2016comment,modanese2018time}. It is important in our opinion to develop a formalism that allows to compute the electromagnetic field generated by local currents also in those cases.

In this work, after recalling recent progress in the theory and numerical solutions of the extended Maxwell-Aharonov-Bohm equations, we derive in Sect.\ \ref{general} the corresponding wave equations for the electric and magnetic field. These equations allow to compute $\mathbf{E}$ and $\mathbf{B}$ directly from the physical sources $\rho$ and $\mathbf{j}$, without any reference to the scalar field $S$ which in the traditional formulation has the role to restore local conservation through an additional (or ``secondary``) charge density proportional to $\partial_t \rho$ and an additional current density proportional to $\nabla S$. The general solution for localized sources is written in the form of retarded integrals. In Sect.\ \ref{radiative} we compute for the first time the electric and magnetic dipole radiation in far-field approximation. We point out that in general a longitudinal electric radiation field $\mathbf{E}_L$ can be present; this is one of the main predictions of the extended theory for the case of an oscillating current that is not locally conserved. On the other hand we note that the magnetic field $\mathbf{B}$ is simply proportional to the curl $\nabla \times \mathbf{j}$, like in Maxwell's theory. Therefore in the limit of stationary currents, in the anomalous case in which $\nabla \cdot \mathbf{j}$ is not zero everywhere (presence of charge sinks/sources), $\mathbf{B}$ is insensitive to the secondary current $\nabla S$, but can reveal the discontinuities in the current simply because to the ``missing links`` of current correspond missing contributions to $\mathbf{B}$ in the Biot-Savart formula. Such effects are expected to be small, but detectable with accurate experiments, as briefly discussed in Sect.\  \ref{signatures}. Finally, in Sect.\ \ref{numerical} we present some numerical solutions for the case of stationary currents, obtained not through the direct wave equations derived in this work, but through the double-retarded integrals written in \cite{modanese2017electromagnetic,modanese2019design}. In this way it is possible to display explicitly the contributions of the auxiliary field $S$, confirming that such contributions cancel out and the only consequence for $\mathbf{B}$ is the missing field effect.

Our final message can be summarized as follows: microscopic models for the computation of the current density in molecular devices can and should continue to improve their performance and their precision without worrying about local conservation of $\mathbf{j}$. The extended Maxwell equations allow in any case a reliable and efficient computation of the resulting fields. This investigation may lead to interesting discoveries in those cases where the e.m.\ fields generated by molecular currents are strong enough to play a significant role.

\section{General equations for $\mathbf{E}$ and $\mathbf{B}$ and their radiative solution}
\label{general}

In this section we write the general wave equations for the electric and magnetic field in extended electrodynamics, in a form in which the auxiliary field $S$ is completely eliminated, and we find their radiative solution.

The extended Maxwell equations in the Aharonov-Bohm theory are \cite{Modanese2017MPLB, modanese2017electromagnetic}
\begin{eqnarray}
\nabla \cdot \mathbf{E} &=&4\pi \rho-\frac{1}{c}\frac{\partial S}{\partial t} \label{mme-E}\\
\nabla \times \mathbf{E}&=&-\frac{1}{c} \frac{\partial \mathbf{B}}{\partial t} \\
\nabla \cdot \mathbf{B}&=&0 \\
\nabla \times \mathbf{B} &=&\frac{4\pi}{c} \mathbf{j}+\frac{1}{c}\frac{\partial \mathbf{E}}{\partial t}+\nabla S \label{mme-B}
\end{eqnarray}
where CGS units have been employed and $S$ is an auxiliary field whose source is the extra-current $I=\partial_t \rho+\nabla \cdot \mathbf{j}$, namely
\begin{equation}
    \frac{1}{c^{2}}\frac{\partial ^{2}S}{\partial t^{2}}-\nabla ^{2}S =\frac{4\pi}{c}
\left[ \frac{\partial \rho }{\partial t}+\nabla \cdot \mathbf{j}\right]
\equiv \frac{4\pi}{c}I
\label{eq-S}
\end{equation}
The field $S$ is clearly zero in the pure Maxwell theory, which requires strict local conservation of the current.

It is possible to interpret the extended equations (\ref{mme-E}), (\ref{mme-B}), (\ref{eq-S}) as involving some additional or ``secondary`` sources, namely a secondary charge density proportional to $\partial_t S$ and a secondary current density proportional to $\nabla S$. Including these additional charge and current densities gives total densities which satisfy the continuity equation. Note that the solution of (\ref{eq-S}) for a localized extra-source yields $S$ as a retarded integral:
\begin{equation*}
S\left( \mathbf{x},t\right) =\frac{1}{c}\int \frac{I\left( 
\mathbf{x}^{\prime },t^{\prime }\right) }{\left\vert \mathbf{x}-\mathbf{x}
^{\prime }\right\vert }d^{3}x^{\prime },
\end{equation*}
with $t^{\prime }=t-\left\vert \mathbf{x}-\mathbf{x}^{\prime }\right\vert /c$.
Therefore $S$ is not localized in the region where the physical sources are present. For this reason we have called the secondary charge and current in our previous work ``cloud charge'' and ``cloud current'' and we have evaluated them in some specific cases (see also the numerical simulations in Sect.\ \ref{numerical} of this work).

The secondary current, in particular, coincides with the additional current predicted by the Landauer-B\"uttiker theory for systems with quantum transport in which local conservation of the current fails \cite{li2008definition,zhang2011first}.

It is possible, however, to write wave equations for $\mathbf{E}$ and $\mathbf{B}$ in which the field $S$ and the secondary charge and current are completely absent. This is in some sense reassuring, because it implies that the physical fields only depend on the localized, physical sources, and that there is no reason to regard the secondary currents as real and to care, for instance, about their dissipation properties. (See also an alternative proof of the independence of the electric and magnetic fields on the scalar source in Appendix \ref{alternative}.)

Although the extended Aharonov-Bohm theory has only a limited gauge invariance, it is possible to define potentials $\phi$ and $\mathbf{A}$ with the usual relations to $\mathbf{E}$ and $\mathbf{B}$, namely
\begin{eqnarray*}
\mathbf{E} &=&-\nabla \phi -\frac{1}{c} \frac{\partial \mathbf{A}}{\partial t}, \\
\mathbf{B} &=&\nabla \times \mathbf{A}.
\end{eqnarray*}
and to write their wave equations as

\begin{eqnarray}
\frac{1}{c^{2}}\frac{\partial ^{2}\phi }{\partial t^{2}}-\nabla ^{2}\phi &=&%
4\pi \rho -\frac{1}{c}\frac{\partial S}{\partial t}, \label{wave-phi}\\
\frac{1}{c^{2}}\frac{\partial ^{2}\mathbf{A}}{\partial t^{2}}-\nabla ^{2}%
\mathbf{A} &=&\frac{4\pi}{c}\mathbf{j}+\nabla S \label{wave-A}
\end{eqnarray}
where once again the role of the secondary charge and current is evident.
The solutions of (\ref{wave-phi}) and (\ref{wave-A}) are
\begin{eqnarray*}
\phi \left( \mathbf{x},t\right) &=&\int \frac{\rho \left( \mathbf{x}^{\prime
},t^{\prime }\right) }{\left\vert \mathbf{x}-\mathbf{x}^{\prime }\right\vert 
}d^{3}x^{\prime }-\frac{1}{4\pi c}\int \frac{1}{\left\vert \mathbf{x}-%
\mathbf{x}^{\prime }\right\vert }\frac{\partial }{\partial t^{\prime }}%
S\left( \mathbf{x}^{\prime },t^{\prime }\right) d^{3}x^{\prime }, \\
\mathbf{A}\left( \mathbf{x},t\right) &=&\frac{1}{c}\int \frac{\mathbf{j}%
\left( \mathbf{x}^{\prime },t^{\prime }\right) }{\left\vert \mathbf{x}-%
\mathbf{x}^{\prime }\right\vert }d^{3}x^{\prime }+\frac{1}{4\pi }\int \frac{%
\nabla ^{\prime }S\left( \mathbf{x}^{\prime },t^{\prime }\right) }{%
\left\vert \mathbf{x}-\mathbf{x}^{\prime }\right\vert }d^{3}x^{\prime }.
\end{eqnarray*}

By taking the gradient of (\ref{wave-phi}) and adding the time derivative of (\ref{wave-A}) one obtains an equation for $\mathbf{E}$ without $S$, while taking the curl of (\ref{wave-A}) one obtains an equation for $\mathbf{B}$ without $S$:

\begin{eqnarray}
\frac{1}{c^{2}}\frac{\partial ^{2}\mathbf{E}}{\partial t^{2}}-\nabla ^{2}%
\mathbf{E} &=&=-\frac{4\pi}{c^2} \left( \frac{\partial \mathbf{j}}{\partial
t}+c^{2}\nabla \rho \right) , \label{wave-E}\\
\frac{1}{c^{2}}\frac{\partial ^{2}\mathbf{B}}{\partial t^{2}}-\nabla ^{2}%
\mathbf{B} &=&\frac{4\pi}{c}\nabla \times \mathbf{j}. \label{wave-B}
\end{eqnarray}

These equations have also been derived in \cite{hively2019classical}, starting from the extended field equations (\ref{mme-E}), (\ref{mme-B}), (\ref{eq-S}) and using some vector calculus identities.

For localized sources the corresponding solutions are%
\begin{eqnarray}
\mathbf{E}\left( \mathbf{x},t\right) &=&-\frac{1}{c^2}\int \left( 
\frac{\partial \mathbf{j}}{\partial t^{\prime }}+c^{2}\nabla ^{\prime }\rho
\right) \frac{d^{3}x^{\prime }}{\left\vert \mathbf{x}-\mathbf{x}^{\prime
}\right\vert }, \label{solE}\\
\mathbf{B}\left( \mathbf{x},t\right) &=&\frac{1}{c}\int \nabla
^{\prime }\times \mathbf{j}\frac{d^{3}x^{\prime }}{\left\vert \mathbf{x}-%
\mathbf{x}^{\prime }\right\vert }. \label{solB}
\end{eqnarray}

We recall that the current density $\mathbf{j}$ can be in general written as the sum of an ``irrotational'' part, which is the gradient of a scalar field and has zero curl, plus a ``solenoidal'' part, which is the curl of a vector field and has zero divergence:
\begin{equation}
    \mathbf{j}=\mathbf{j}_{irrot}+\mathbf{j}_{solen} \equiv \nabla F + \nabla \times \mathbf{G}
\end{equation}
According to eq.\ (\ref{solB}), the irrotational part has no influence on the magnetic field. It follows in particular that interruptions in the current inevitably cause a ``missing field'' effect, as discussed in Sects.\ \ref{signatures}, \ref{numerical}. (Because the secondary current $\nabla S$ which restores current conservation in the extended equations is purely irrotational.) On the other hand, any change in the solenoidal part (like e.g.\ in \cite{dreyer2018current}) does not affect current conservation but is reflected in a change in the field.

One can further operate on the expressions (\ref{solE}), (\ref{solB}) considering that%
\begin{equation*}
\int \frac{\partial \mathbf{j}}{\partial t^{\prime }}\frac{d^{3}x^{\prime }}{%
\left\vert \mathbf{x}-\mathbf{x}^{\prime }\right\vert }=\int \frac{\partial 
\mathbf{j}}{\partial t}\frac{d^{3}x^{\prime }}{\left\vert \mathbf{x}-\mathbf{%
x}^{\prime }\right\vert }=\frac{\partial }{\partial t}\int \frac{\mathbf{j}%
\left( \mathbf{x}^{\prime },t^{\prime }\right) }{\left\vert \mathbf{x}-%
\mathbf{x}^{\prime }\right\vert }d^{3}x^{\prime },
\end{equation*}
and that%
\begin{eqnarray*}
\int \frac{\nabla ^{\prime }\rho }{\left\vert \mathbf{x}-\mathbf{x}^{\prime
}\right\vert }d^{3}x^{\prime } &=&\int \nabla ^{\prime }\left[ \frac{\rho
\left( \mathbf{x}^{\prime },t^{\prime }\right) }{\left\vert \mathbf{x}-%
\mathbf{x}^{\prime }\right\vert }\right] d^{3}x^{\prime } \\
&&-\int \rho \left( \mathbf{x}^{\prime },t^{\prime }\right) \nabla ^{\prime }%
\frac{1}{\left\vert \mathbf{x}-\mathbf{x}^{\prime }\right\vert }%
d^{3}x^{\prime } \\
&=&\oint \frac{\rho \left( \mathbf{x}^{\prime },t^{\prime }\right) }{%
\left\vert \mathbf{x}-\mathbf{x}^{\prime }\right\vert }d\mathbf{S}^{\prime
}+\nabla \frac{\rho \left( \mathbf{x}^{\prime },t^{\prime }\right) }{%
\left\vert \mathbf{x}-\mathbf{x}^{\prime }\right\vert }d^{3}x^{\prime } \\
&=&\nabla \int \frac{\rho \left( \mathbf{x}^{\prime },t^{\prime }\right) }{%
\left\vert \mathbf{x}-\mathbf{x}^{\prime }\right\vert }d^{3}x^{\prime },
\end{eqnarray*}%
where the surface integral is zero because there is no charge on the
external surface of the volume of the source, and the change from $\nabla
^{\prime }$ to $-\nabla $ in the integral in the second line was done
because the function on which it operates is of argument $\mathbf{x}-\mathbf{%
x}^{\prime }$.

We have in a similar manner%
\begin{eqnarray*}
\int \frac{\nabla ^{\prime }\times \mathbf{j}}{\left\vert \mathbf{x}-\mathbf{%
x}^{\prime }\right\vert }d^{3}x^{\prime } &=&\int \nabla ^{\prime }\times %
\left[ \frac{\mathbf{j}\left( \mathbf{x}^{\prime },t^{\prime }\right) }{%
\left\vert \mathbf{x}-\mathbf{x}^{\prime }\right\vert }\right]
d^{3}x^{\prime } \\
&&-\int \nabla ^{\prime }\left( \frac{1}{\left\vert \mathbf{x}-\mathbf{x}%
^{\prime }\right\vert }\right) \times \mathbf{j}\left( \mathbf{x}^{\prime
},t^{\prime }\right) d^{3}x^{\prime } \\
&=&-\oint \frac{\mathbf{j}\left( \mathbf{x}^{\prime },t^{\prime }\right) }{%
\left\vert \mathbf{x}-\mathbf{x}^{\prime }\right\vert }\times d\mathbf{S}%
^{\prime } \\
&&+\int \nabla \left( \frac{1}{\left\vert \mathbf{x}-\mathbf{x}^{\prime
}\right\vert }\right) \times \mathbf{j}\left( \mathbf{x}^{\prime },t^{\prime
}\right) d^{3}x^{\prime } \\
&=&\nabla \times \int \frac{\mathbf{j}\left( \mathbf{x}^{\prime },t^{\prime
}\right) }{\left\vert \mathbf{x}-\mathbf{x}^{\prime }\right\vert }%
d^{3}x^{\prime }.
\end{eqnarray*}

The expressions of the fields thus reduce to 
\begin{subequations}
\label{emhat}
\begin{eqnarray}
\mathbf{E}\left( \mathbf{x},t\right) &=&-\frac{1}{c^{2}}\frac{\partial }{%
\partial t}\int \frac{\mathbf{j}\left( \mathbf{x}^{\prime },t^{\prime
}\right) }{\left\vert \mathbf{x}-\mathbf{x}^{\prime }\right\vert }%
d^{3}x^{\prime }-\nabla \int \frac{\rho \left( \mathbf{x}^{\prime
},t^{\prime }\right) }{\left\vert \mathbf{x}-\mathbf{x}^{\prime }\right\vert 
}d^{3}x^{\prime },  \label{ehat} \\
\mathbf{B}\left( \mathbf{x},t\right) &=&\frac{1}{c}\nabla \times \int \frac{%
\mathbf{j}\left( \mathbf{x}^{\prime },t^{\prime }\right) }{\left\vert 
\mathbf{x}-\mathbf{x}^{\prime }\right\vert }d^{3}x^{\prime }.  \label{bhat}
\end{eqnarray}

These expressions show that the EM fields are obtained from the potentials
evaluated without considering their source terms depending on $S$.

\subsection{Radiative solution in the dipole approximation}
\label{radiative}

For a pure temporal Fourier mode: $\mathbf{j}\left( \mathbf{x},t\right) =%
\widehat{\mathbf{j}}\left( \mathbf{x}\right) \exp \left( -i\omega t\right) $%
, $\rho \left( \mathbf{x},t\right) =\widehat{\rho }\left( \mathbf{x}\right)
\exp \left( -i\omega t\right) $, the general solution for the corresponding
scalar potential, without the source depending on $S$, is of the form $\phi
\left( \mathbf{x},t\right) =\widehat{\phi }\left( \mathbf{x}\right) \exp
\left( -i\omega t\right) $, with 
\end{subequations}
\begin{equation*}
\widehat{\phi }\left( \mathbf{x}\right) =\int \frac{\widehat{\rho }\left( 
\mathbf{x}^{\prime }\right) }{\left\vert \mathbf{x}-\mathbf{x}^{\prime
}\right\vert }\exp \left( ik\left\vert \mathbf{x}-\mathbf{x}^{\prime
}\right\vert \right) d^{3}x^{\prime },
\end{equation*}%
where $k=\omega /c$. For a distant observation point and wavelength large as
compared with the source dimension, the expression is approximated by%
\begin{equation*}
\widehat{\phi }\left( \mathbf{x}\right) =\frac{\exp \left( ikr\right) }{r}%
\int \widehat{\rho }\left( \mathbf{x}^{\prime }\right) \left( 1-ik\mathbf{n}%
\cdot \mathbf{x}^{\prime }\right) d^{3}x^{\prime },
\end{equation*}%
with \ $r=\left\vert \mathbf{x}\right\vert $, and $\mathbf{n}=\mathbf{x}/r$.

Even if the charge is not conserved locally, it is conserved globally, so
that%
\begin{equation*}
\int \widehat{\rho }\left( \mathbf{x}^{\prime }\right) d^{3}x^{\prime }=0,
\end{equation*}%
and thus%
\begin{eqnarray*}
\widehat{\phi }\left( \mathbf{x}\right) &=&-i\frac{k}{r}\exp \left(
ikr\right) \mathbf{n}\cdot \int \widehat{\rho }\left( \mathbf{x}^{\prime
}\right) \mathbf{x}^{\prime }d^{3}x^{\prime } \\
&\equiv &-i\frac{k}{r}\exp \left( ikr\right) \mathbf{n}\cdot \widehat{%
\mathbf{p}},
\end{eqnarray*}%
where $\widehat{\mathbf{p}}$ is the Fourier amplitude of the usual charge
dipole.

The solution to the corresponding Fourier amplitude of the vector potential
is, with the same approximations,%
\begin{equation*}
\widehat{\mathbf{A}}\left( \mathbf{x}\right) =\frac{1}{cr}\exp \left(
ikr\right) \int \widehat{\mathbf{j}}\left( \mathbf{x}^{\prime }\right)
d^{3}x^{\prime }.
\end{equation*}

Using the identity%
\begin{eqnarray*}
\int \widehat{\mathbf{j}}\left( \mathbf{x}^{\prime }\right) d^{3}x^{\prime
} &=&\oint \mathbf{x}^{\prime }\widehat{\mathbf{j}}\left( \mathbf{x}%
^{\prime }\right) \cdot d\mathbf{S}^{\prime }-\int \mathbf{x}^{\prime
}\nabla ^{\prime }\cdot \widehat{\mathbf{j}}\left( \mathbf{x}^{\prime
}\right) d^{3}x^{\prime } \\
&=&-\int \mathbf{x}^{\prime }\nabla ^{\prime }\cdot \widehat{\mathbf{j}}%
\left( \mathbf{x}^{\prime }\right) d^{3}x^{\prime },
\end{eqnarray*}%
where in the second line it was used that there is no current leaving or
entering the volume of the source, and also allowing for charge
non-conservation, so that%
\begin{equation*}
\nabla ^{\prime }\cdot \widehat{\mathbf{j}}\left( \mathbf{x}^{\prime
}\right) =i\omega \widehat{\rho }\left( \mathbf{x}^{\prime }\right) +%
\widehat{I}\left( \mathbf{x}^{\prime }\right) ,
\end{equation*}%
we have%
\begin{eqnarray*}
\widehat{\mathbf{A}}\left( \mathbf{x}\right) &=&-i\frac{\omega }{cr}\exp
\left( ikr\right) \int \mathbf{x}^{\prime }\widehat{\rho }\left( \mathbf{x}%
^{\prime }\right) d^{3}x^{\prime }-\frac{1}{cr}\exp \left( ikr\right) \int 
\mathbf{x}^{\prime }\widehat{I}\left( \mathbf{x}^{\prime }\right)
d^{3}x^{\prime } \\
&=&-i\frac{\omega }{cr}\exp \left( ikr\right) \widehat{\mathbf{p}}-\frac{1}{%
cr}\exp \left( ikr\right) \widehat{\mathbf{P}},
\end{eqnarray*}%
where the dipolar moment of the Fourier amplitude of the extra source $I$
was defined.%
\begin{equation*}
\widehat{\mathbf{P}}=\int \mathbf{x}^{\prime }\widehat{I}\left( \mathbf{x}%
^{\prime }\right) d^{3}x^{\prime }.
\end{equation*}

The Fourier amplitude of the radiative electric field is thus%
\begin{eqnarray}
\widehat{\mathbf{E}}\left( \mathbf{x}\right)  &=&\frac{i\omega }{c}\widehat{%
\mathbf{A}}\left( \mathbf{x}\right) -\nabla \widehat{\phi }\left( \mathbf{x}%
\right)  \\
&=&\frac{\omega ^{2}}{c^{2}r}\exp \left( ikr\right) \left[ \widehat{\mathbf{p%
}}-\left( \mathbf{n}\cdot \widehat{\mathbf{p}}\right) \mathbf{n}\right] -i%
\frac{\omega }{c^{2}r}\exp \left( ikr\right) \widehat{\mathbf{P}} \\
&=&-\frac{\omega ^{2}}{c^{2}r}\exp \left( ikr\right) \left[ \left( \widehat{%
\mathbf{p}}\times \mathbf{n}\right) \times \mathbf{n}\right] -i\frac{\omega 
}{c^{2}r}\exp \left( ikr\right) \widehat{\mathbf{P}}. \label{E-Fourier}
\end{eqnarray}

Analogously,%
\begin{equation*}
\widehat{\mathbf{B}}\left( \mathbf{x}\right) =\nabla \times \widehat{\mathbf{%
A}}\left( \mathbf{x}\right) =\frac{1}{c^{2}r}\exp \left( ikr\right) \left(
-\omega ^{2}\widehat{\mathbf{p}}+i\omega \widehat{\mathbf{P}}\right) \times 
\mathbf{n}.
\end{equation*}

Transforming back to the time domain the EM fields are given by 
\begin{eqnarray}
\mathbf{E}\left( \mathbf{x},t\right)  &=&-\sum_{\omega }\frac{\omega ^{2}}{%
c^{2}r}\exp \left[ -i\omega \left( t-r/c\right) \right] \left\{ \left[ 
\widehat{\mathbf{p}}\left( \omega \right) \times \mathbf{n}\right] \times 
\mathbf{n}\right\}  \\
&&-\sum_{\omega }i\frac{\omega }{c^{2}r}\exp \left[ -i\omega \left(
t-r/c\right) \right] \widehat{\mathbf{P}}\left( \omega \right)  \\
&=&\frac{1}{c^{2}r}\frac{\partial ^{2}}{\partial t^{2}}\sum_{\omega }\exp %
\left[ -i\omega \left( t-r/c\right) \right] \left\{ \left[ \widehat{\mathbf{p%
}}\left( \omega \right) \times \mathbf{n}\right] \times \mathbf{n}\right\} 
\\
&&+\frac{1}{c^{2}r}\frac{\partial }{\partial t}\sum_{\omega }\exp \left[
-i\omega \left( t-r/c\right) \right] \widehat{\mathbf{P}}\left( \omega
\right)  \\
&=&\frac{1}{c^{2}r}\left\{ \left[ \overset{..}{\mathbf{p}}\left(
t-r/c\right) \times \mathbf{n}\right] \times \mathbf{n}+\overset{.}{\mathbf{P%
}}\left( t-r/c\right) \right\} , \label{E-time}
\end{eqnarray}%
and 
\begin{eqnarray*}
\mathbf{B}\left( \mathbf{x},t\right)  &=&\frac{1}{c^{2}r}\sum_{\omega }\exp %
\left[ -i\omega \left( t-r/c\right) \right] \left[ -\omega ^{2}\widehat{%
\mathbf{p}}\left( \omega \right) +i\omega \widehat{\mathbf{P}}\left( \omega
\right) \right] \times \mathbf{n} \\
&=&\frac{1}{c^{2}r}\frac{\partial ^{2}}{\partial t^{2}}\sum_{\omega }\exp %
\left[ -i\omega \left( t-r/c\right) \right] \widehat{\mathbf{p}}\left(
\omega \right) \times \mathbf{n} \\
&&-\frac{1}{c^{2}r}\frac{\partial }{\partial t}\sum_{\omega }\exp \left[
-i\omega \left( t-r/c\right) \right] \widehat{\mathbf{P}}\left( \omega
\right) \times \mathbf{n} \\
&=&\frac{1}{c^{2}r}\left[ \overset{..}{\mathbf{p}}\left( t-r/c\right) -%
\overset{.}{\mathbf{P}}\left( t-r/c\right) \right] \times \mathbf{n}.
\end{eqnarray*}

The first term of $\mathbf{E}$ vanishes when it is multiplied by $\mathbf{n}$ and therefore represents the transverse component. The longitudinal component is just $\mathbf{E}\cdot \mathbf{n}$, i.e.
\begin{equation}
    \mathbf{E}_L= \frac{1}{c^2 r} \Dot{\mathbf{P}}(t-r/c)\cdot \mathbf{n}
    \label{Elong}
\end{equation}
To fix the ideas, suppose that the moment $\widehat{\mathbf{P}}$ of the extra-current is directed along the $z$-axis. Then the component $E_L$ at a fixed distance $r$ is seen to be maximum on the $z$-axis and zero in the $x$-$y$ plane. The opposite happens with the transverse component.

A simple formal example of oscillating extra-current has been introduced in \cite{modanese2019high}. Consider a point-like charge $q$ which oscillates between the positions $(0,0,-a)$ and $(0,0,a)$, without a corresponding current:
\begin{equation}
    \rho(\mathbf{x},t)=q\cos(\omega t) [\delta^3(\mathbf{x}+\mathbf{a})-\delta^3(\mathbf{x}-\mathbf{a})]
\end{equation}
The moment $\widehat{\mathbf{P}}$ in this case has the only non-zero component $|\hat{P}_z|=2\omega q a$. If we assign the physical parameters $q$, $a$ and $\omega$ it is straightforward to compute the longitudinal far field and we have along the $z$-axis $|\hat{E}_L|=2\omega^2qa/(c^2r)$ (in SI units: $|\hat{E}_L|=\mu_0\omega^2qa/(2\pi r)$).  The transverse far field is vanishing.  

More realistically we can suppose that if a local violation of charge conservation occurs, this will only concern a small fraction $\eta$ of the oscillating charge, while the rest of the charge will have a corresponding current and consequently will not generate any longitudinal radiation field. In that case, the ratio between $E_L$ and $E_T$ will approximately be equal to $\eta$. 

In \cite{modanese2019high} we performed a numerical simulation of a source of this kind with $a=10^{-7}$ cm, $\omega=2\pi \cdot 10^9$ Hz, $\theta=\pi/4$ ($\theta$ is the 3D polar coordinate), and we obtained that for a totally anomalous source (i.e. $\eta=1$, the entire charge oscillates without a current) the longitudinal field is much larger that the transverse field that would be generated by a corresponding regular source at the same position. The field was computed at a relatively small distance ($r$ varied between $3\lambda$ and $13\lambda$), therefore the result cannot be directly compared with eq.\ (\ref{Elong}), which holds for larger distances; still it confirms the presence of a longitudinal field, because even in the near-field range it is impossible to obtain such big longitudinal components if the source satisfies local conservation.

In conclusion, if oscillating currents exist that violate even partially the continuity condition, a sizable longitudinal electric radiation should be generated which is obviously incompatible with the standard Maxwell equations.

\section{Possible experimental signatures of a missing $\mathbf{B}$ and a radiative $\mathbf{E}_L$}
\label{signatures}

Let us first consider a possible experimental observation of the missing field effect for stationary currents. As discussed in \cite{modanese2019design} and confirmed in Sects.\ \ref{general} and \ref{signatures} of this work, if in a linear conductor there are regions in which a violation of continuity occurs and a fraction $\eta$ of the current does not flow as ``physical current $\rho \mathbf{v}$'' but as ``secondary current $\nabla S$'', then the magnetic field generated by the secondary current is zero, as a consequence of eq.\ (\ref{wave-B}). To detect the missing field, we proposed in \cite{modanese2019design} to measure at a fixed distance $r$ the field of a normal conductor carrying a current $i$, and then the field of the anomalous conductor carrying the same current. A differential measurement with three wires was devised, in order to minimize errors. The scheme was especially suited for brief current pulses and for the case when the supposed anomalous conductor is a superconductor, which can be driven in/out a normal state by changing its temperature.

Another possible technique, which employs stationary currents and one single wire, is based on the measurement of the ratio $B/i$ at a fixed distance in dependence on $i$ (supposing that a variable external bias allows to change $i$). 
This would work if the ratio $\eta$ depends on $i$, in which case $B/i$ is expected to change, in open violation of the Maxwell equations. In fact, the current patterns and local discontinuities observed in simulations of molecular devices depend in general on the total current.

Possible errors could originate from slight changes in the spatial distribution of the current at different $i$, if the conductor has a radius that cannot be disregarded compared to the distance $r$. This radius depends on how ``elementary'' the conductor is (for example, a single carbon nanowire vs.\ a bundle of nanowires); in turn, that depends on how much current is needed for the measurement, and thus indirectly on $r$ (because of the size and sensitivity of the detector).

For example, suppose that a single carbon nanowire can carry a current of $10^{-10}$ A and the detector can be placed at a distance of $10^{-4}$ m, very large compared to the radius of the wire. The field would then be of the order of $10^{-12}$ T, i.e.\ accessible to a sensitive SQUID. If the ratio $\eta$ is of the order of $10^{-2}$ and the error on the current is negligible, the SQUID is required to detect a variation in the field of $10^{-14}$ T.

Turning now to electric fields, the possible generation and detection of a longitudinal e.m.\ radiation has rarely been explored in the past decades \cite{giakos1993detection,monstein2002observation,monstein2004remarks,butterworth2013longitudinal,umul2018excitation,simulik2019slightly}. A recent preliminary experiment and its relation to the extended electromagnetic theory has been described in \cite{hively2019classical}, including a discussion of error sources. A distinctive feature of longitudinal electric radiation would be its ability to penetrate thin layers of good conductors much easier than a transverse radiation. Even in favourable cases, however, a predominant transverse component would be present, causing interference and noise. If future developments can lead to a clean selective detection of the longitudinal component, the possible technological applications would be manifold.

There is also much to do, of course, concerning the design of efficient antennas. First, one would need to identify materials with local non-conservation that can support a sufficiently large current at high frequency. Then the most appropriate antenna geometry should be studied. For this purpose, the general equations developed in this paper constitute a firm starting point.

\section{Numerical solutions of the extended equations with the potentials}
\label{numerical}

Consider again the equations for the potentials, written in the form
\begin{align}
	\frac{1}{c^2}\frac{\partial^2 \phi}{\partial t^2}-\nabla^2 \phi=4\pi \rho-\frac{1}{c^2} \frac{\partial}{\partial t} \int d^3y \frac{I\left(t_{ret},\textbf{y} \right)}{\left|\textbf{x}-\textbf{y} \right|}; \label{pot1} \end{align}
\begin{align}
	\frac{1}{c^2}\frac{\partial^2 \textbf{A}}{\partial t^2}-\nabla^2 \textbf{A}=\frac{4\pi}{c} \textbf{j}+\frac{1}{c} \nabla \int d^3y \frac{I\left(t_{ret},\textbf{y} \right)}{\left|\textbf{x}-\textbf{y} \right|}, \label{pot2}
\end{align}
where $I(t,\mathbf{x})$ is an assigned extra-current.
By solving eqs.\ (\ref{pot1}) and (\ref{pot2}) we obtain for $\phi$ and ${\bf A}$ expressions which contain ``double-retarded integrals'' of the extra-current $I$. In \cite{modanese2019design} we gave some numerical computations of these integrals for a slowly varying source, as recalled below.
In this section we consider the stationary case and the resulting expressions are much simpler. For the numerical integration we employ a 6D Monte Carlo, which can be quite time consuming but is straightforward and does not require any analytical approximations. 

The idea is to take, instead of a source with a slow temporal variation, a truly stationary source. Then the retarded integrals become simple space integrals. We recall that the typical time scale of the source in the work \cite{modanese2019design} was $\tau \simeq 10^{-5}$ s, chosen because (1) it corresponds to the proposed experimental conditions, being the characteristic discharge time of the circuit; (2) it allows to disregard certain phenomena occurring only at high frequency, like the intervention of stray capacitance and temporary charge accumulation.

In \cite{modanese2019design} we considered a case of local non-conservation involving a point-like sink, where current partially disappears, and a point-like source where the current reappears. More precisely, the sizes of sink and source were given by a regulator $\varepsilon$ of the order of $10^{-7}$ cm. The regulator was applied after the first 3D integration (analytical) in $d^3z$. Then the second numerical integration was made (in $d^3y$, over the extended secondary cloud current decreasing like $1/|{\bf y}|^3$), using the command \texttt{NIntegrate} of \texttt{Mathematica} and checking the results by comparison with a Monte Carlo integration in 3D.

The disadvantage of that procedure is the complication at the formal-algebraic level. The expressions obtained after the first integration are quite bulky. At the same time, the assumption of point-like sources is limiting, because when the failure of local conservation occurs in a quantum wavefunction, the regions where $\partial_t \rho+\nabla \cdot {\bf j} \neq 0$ are not pointlike but extended, with a shape more similar to that of a couple of disks.

\begin{figure}[h]
  \begin{center}
\includegraphics[width=14.0cm,height=9.2cm]{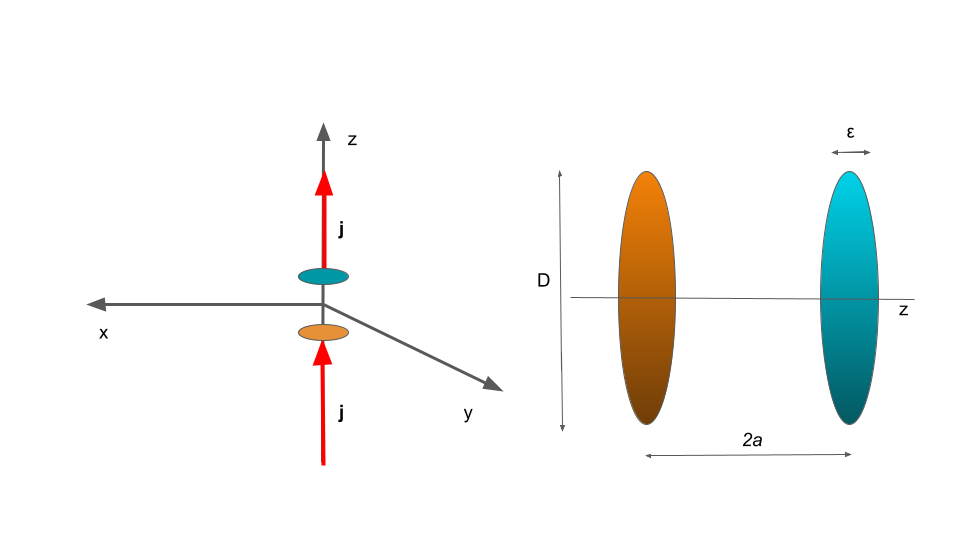}
\caption{Parameters and location of the Gaussian extra-sources employed in the numerical simulations. (Orange: current sink. Blue: current source.)
} 
\label{sources}
  \end{center}
\end{figure}

Therefore we model here the extra-current source with a double Gaussian (Fig.\ \ref{sources}), and the method actually applies to any geometrical shape:
\begin{equation}
I({\bf x})=\frac{I_0}{\sqrt{(2\pi)^3}\varepsilon d^2} \cdot
\end{equation}
\begin{equation}
\left\{ \exp \left[ -\frac{1}{2} \left( \frac{x_1^2}{d^2}+\frac{x_2^2}{d^2}+\frac{(x_3-a)^2}{\varepsilon^2} \right) \right] - \exp \left[ -\frac{1}{2} \left( \frac{x_1^2}{d^2}+\frac{x_2^2}{d^2}+\frac{(x_3+a)^2}{\varepsilon^2} \right) \right] \right\}
\label{gaussian}
\nonumber
\end{equation}

The auxiliary anomalous vector potential written as double integral of the extra-current is
\begin{equation}
{\bf A}({\bf x})=\frac{k}{4\pi} \int d^3y \frac{1}{|{\bf x}-{\bf y}|} \nabla_{\bf y} \int d^3z \frac{I({\bf z})}{|{\bf y}-{\bf z}|}
\label{Astaz}
\end{equation}

When we compute the magnetic field at the position ${\bf x}=(r,0,0)$, its only non-zero component is ${\bf B}_2$:
\begin{equation}
{\bf B}_2({\bf x})=\frac{\partial A_1^s}{\partial x_3}-\frac{\partial A_3^s}{\partial x_1}=
\frac{k}{4\pi} \int d^3y \int d^3z \left( \frac{\partial}{\partial x_3} \frac{1}{|{\bf x}-{\bf y}|} \right) \left( \frac{\partial}{\partial y_1}  \frac{1}{|{\bf y}-{\bf z}|} \right) I({\bf z}) - (1 \leftrightarrow 3)
\end{equation}

The derivative of the first term is
\[
\frac{\partial}{\partial x_3} \frac{1}{\sqrt{(x_1-y_1)^2+(x_2-y_2)^2+(x_3-y_3)^2}} = \frac{y_3-x_3}{\left[ (x_1-y_1)^2+(x_2-y_2)^2+(x_3-y_3)^2 \right]^{3/2}}
\]

The other derivatives are similar, and in total we obtain, also considering that $x_3=0$ in this particular configuration:
\begin{equation}
{\bf B}_2({\bf x})=
\frac{k}{4\pi} \int d^3y \int d^3z \frac{y_3 z_1-x_1 y_3+x_1 z_3-y_1 z_3}{|{\bf x}-{\bf y}|^3 |{\bf y}-{\bf z}|^3} I({\bf z})
\label{B2s}
\end{equation}

This expression is simpler and more direct than those obtained in \cite{modanese2019design} and suitable for a 6D Monte Carlo algorithm which first generates random values of $\mathbf{z}$ in a quite narrow range, corresponding to the support of the Gaussian (\ref{gaussian}), i.e., inside the primary source. The main computational difficulty is (as before, actually) that for ${\bf y}$ it is necessary to sample a much wider region. Typically for each value of ${\bf z}$ we need to generate tens of values of ${\bf y}$ in the sampling. 
 
\begin{figure}[h]
\begin{subfigure}{.5\textwidth}
    \includegraphics[width=7.0cm,height=5.1cm]{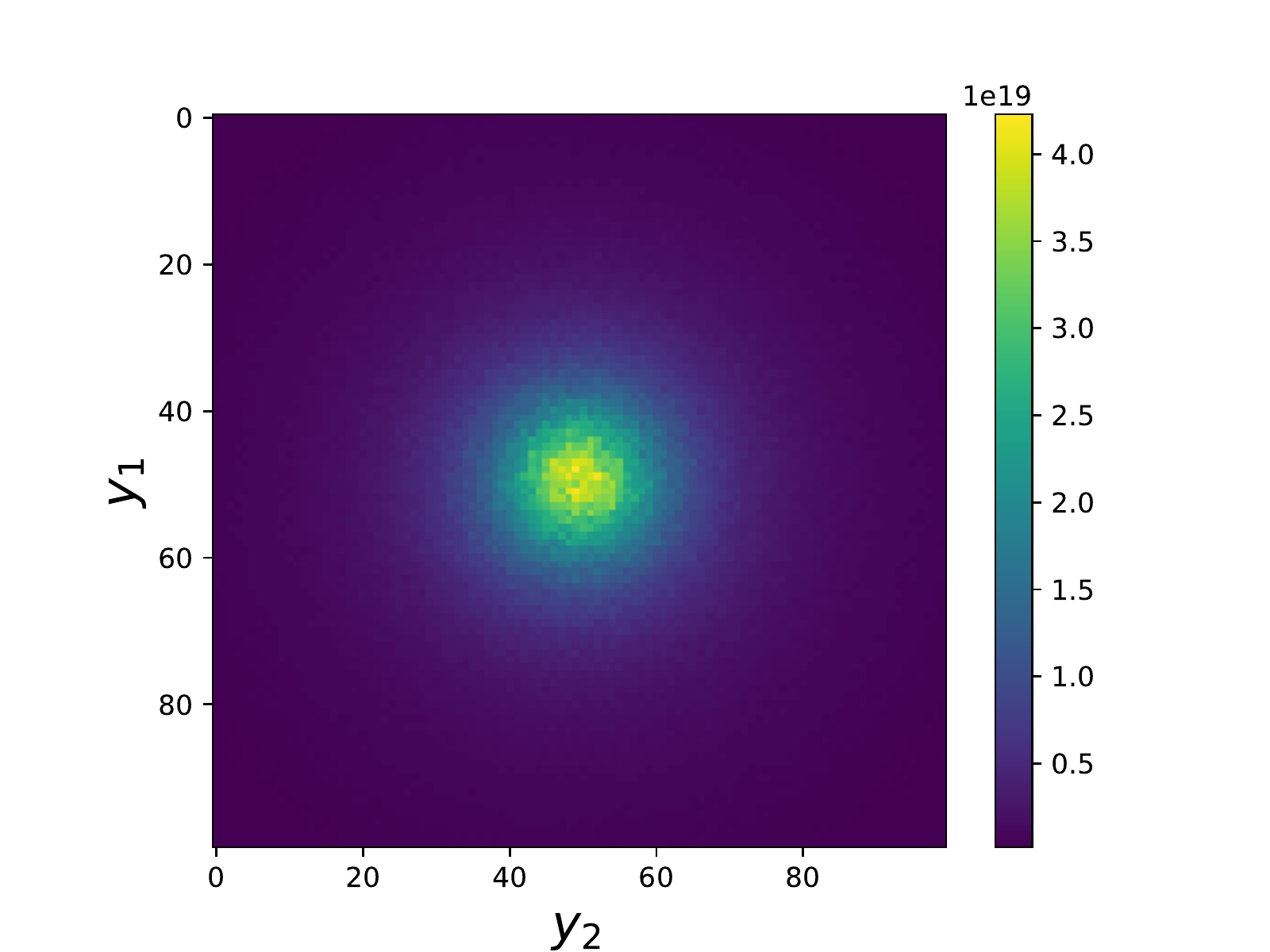}
  \caption{}
\end{subfigure}
\begin{subfigure}{.5\textwidth}
    \includegraphics[width=7.0cm,height=5.1cm]{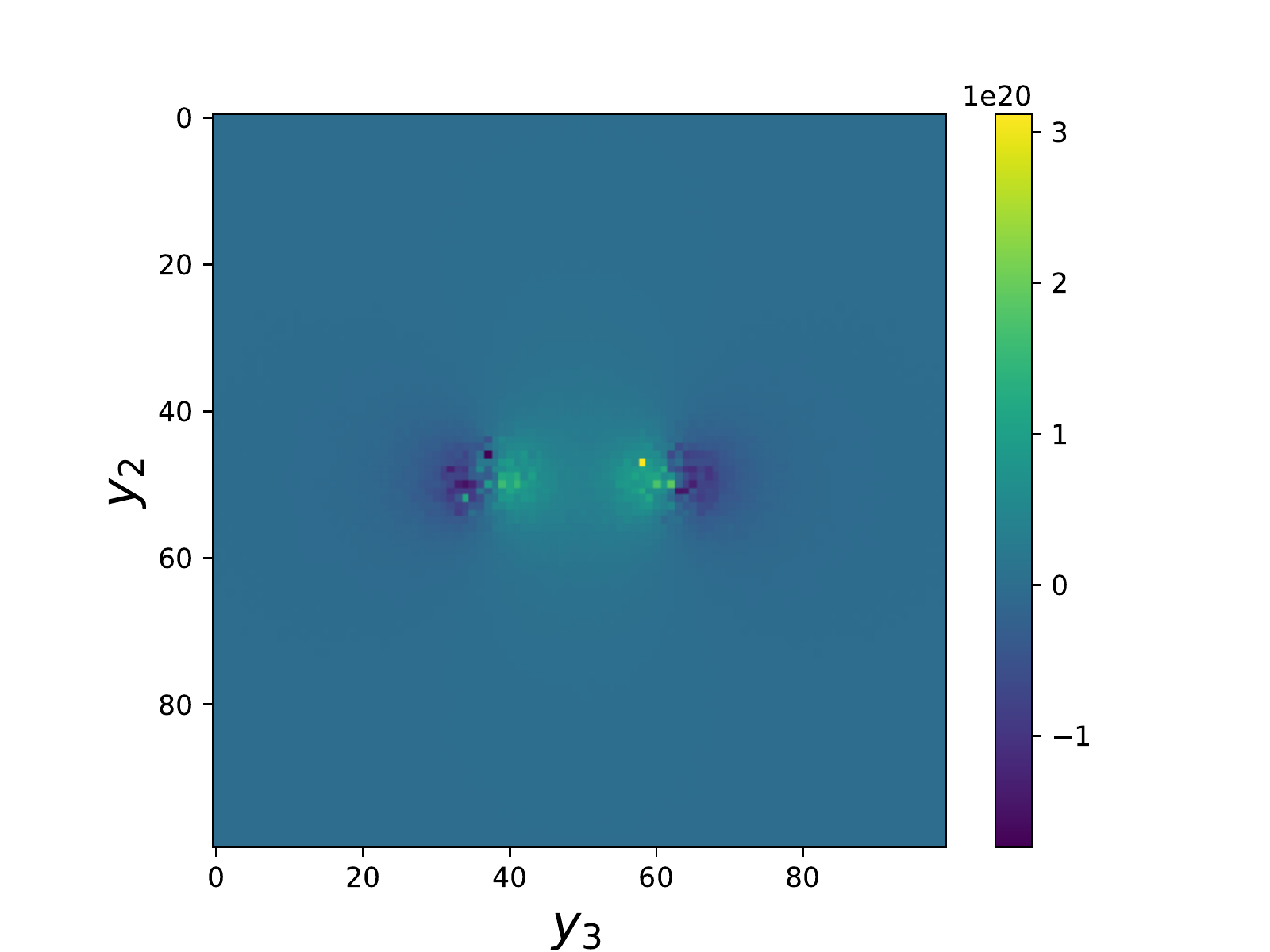}
  \caption{}
\end{subfigure}
\caption{{\bf (a)} Contributions of the ``cloud'' of secondary current to the anomalous field $B^s$ for spherical sink and source (see Tab.\ \ref{table1}), in the $y$ range $[0-10]\cdot 10^{-7}$ cm. In the 6D Monte Carlo integration, the 3D region of the $y$ variable has been subdivided into $100^3$ cells; for each cell, the contribution of sampling points falling into it is displayed. The numerical values on the color scale must be normalized to compute the field, and are therefore not meaningful as absolute value, but their sign and relative values are of interest. In this figure we have $y_3=0$, i.e., we are observing the cloud on the plane equidistant from source and sink. All contributions are positive, since we are between the sources (compare Fig.\ \ref{puntif10-50}). {\bf (b)} Here we have $y_1=0$, therefore we are observing the cloud in the plane that cuts sink and source. Note that sink and source both give positive contributions on the inner side, and negative contributions on the outer side. From Tab.\ \ref{table1} we deduce that the total contribution of the region shown in this figure is positive.
} 
\label{puntif0-10}
\end{figure}

\begin{figure}[h]
  \begin{center}
\includegraphics[width=7.0cm,height=5.1cm]{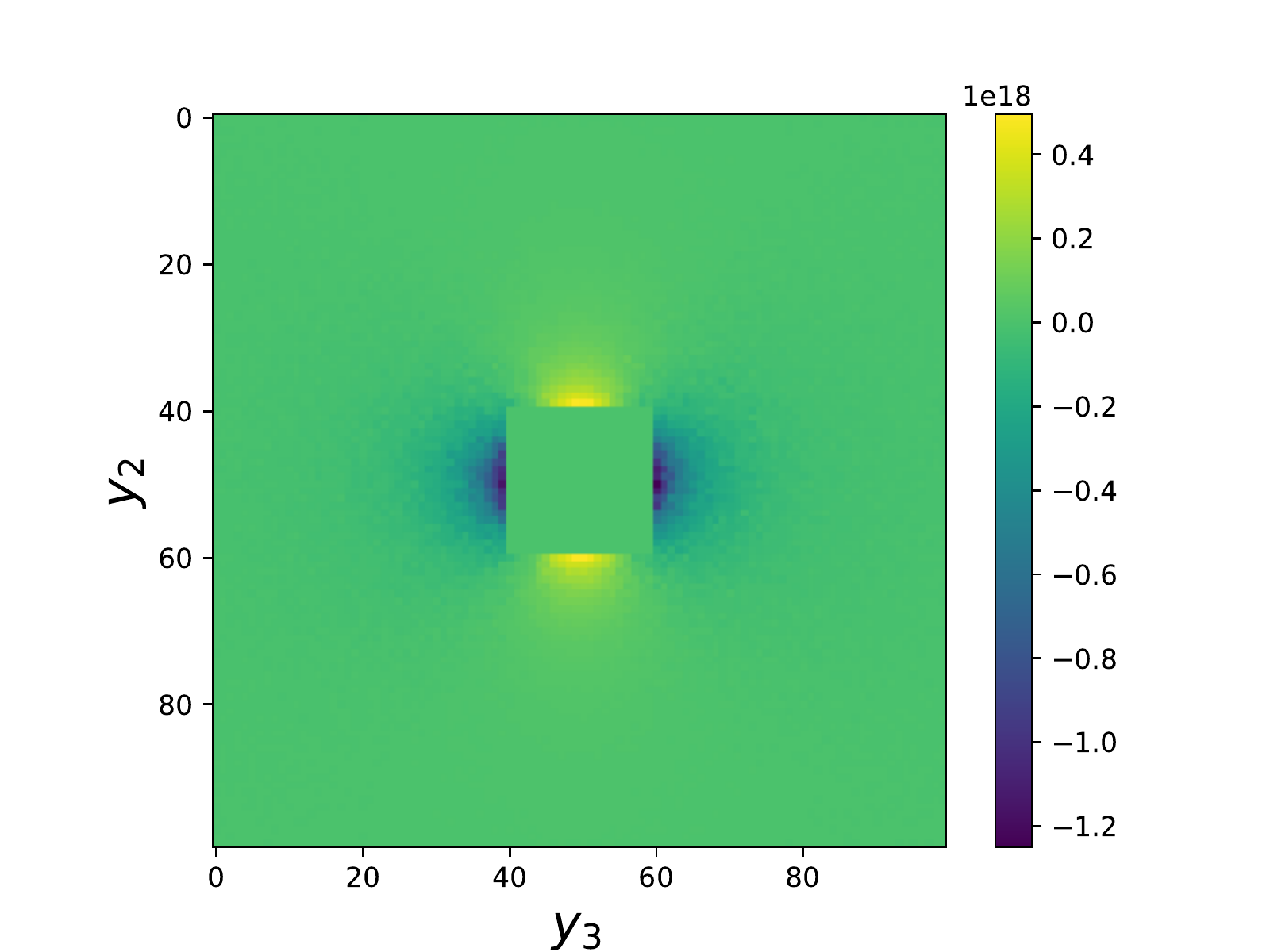}
\caption{Same sources and cutting plane as in Fig.\ \ref{puntif0-10}, (b), but in the $y$ range $[10-50]\cdot 10^{-7}$ cm. (The central region $[0-10]$ is not sampled because it would give a strong noise, compared to the larger region.) Note that the pattern of Fig.\ \ref{puntif0-10}, (b) is confirmed concerning the inner/outer regions with positive/negative contributions respectively.
} 
\label{puntif10-50}
  \end{center}
\end{figure}

\begin{figure}[h]
  \begin{center}
\includegraphics[width=7.0cm,height=5.1cm]{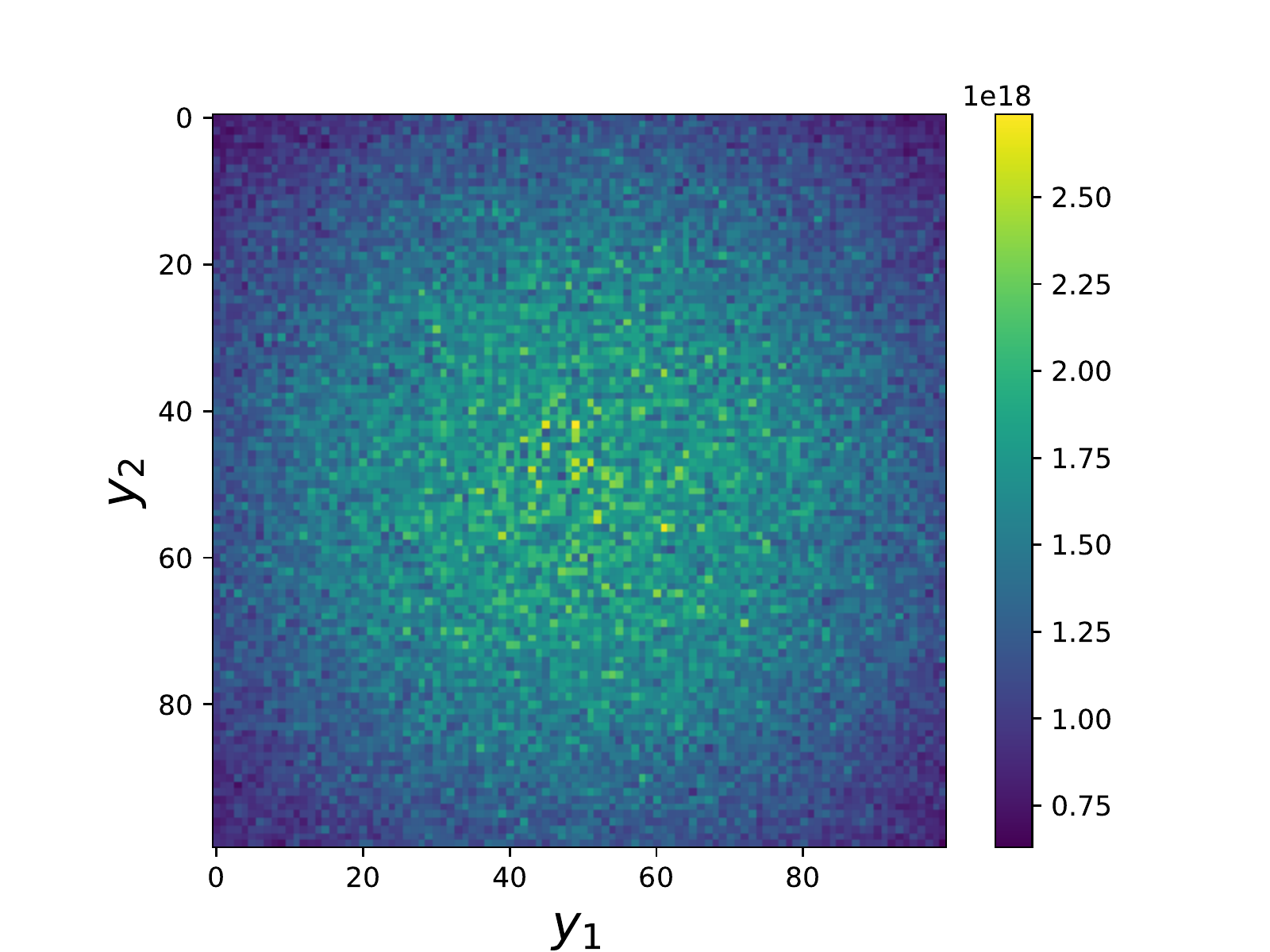}
\caption{This figure can be compared with Fig.\ \ref{puntif0-10}, (a), with the only difference that sink and source are here Gaussian disks/ellypsoids of diameter $D=10\cdot 10^{-7}$ cm instead of $D=0.5\cdot 10^{-7}$ cm. ($D$ corresponds to the $\sigma$ of the Gaussian density distribution.) As in Fig.\ \ref{puntif0-10}, (a), the contributions shown lie on the plane between source and sink, with $y_3=0$.
} 
\label{disco0-10-y3-zero}
  \end{center}
\end{figure}

\begin{figure}[h]
  \begin{center}
\includegraphics[width=7.0cm,height=5.1cm]{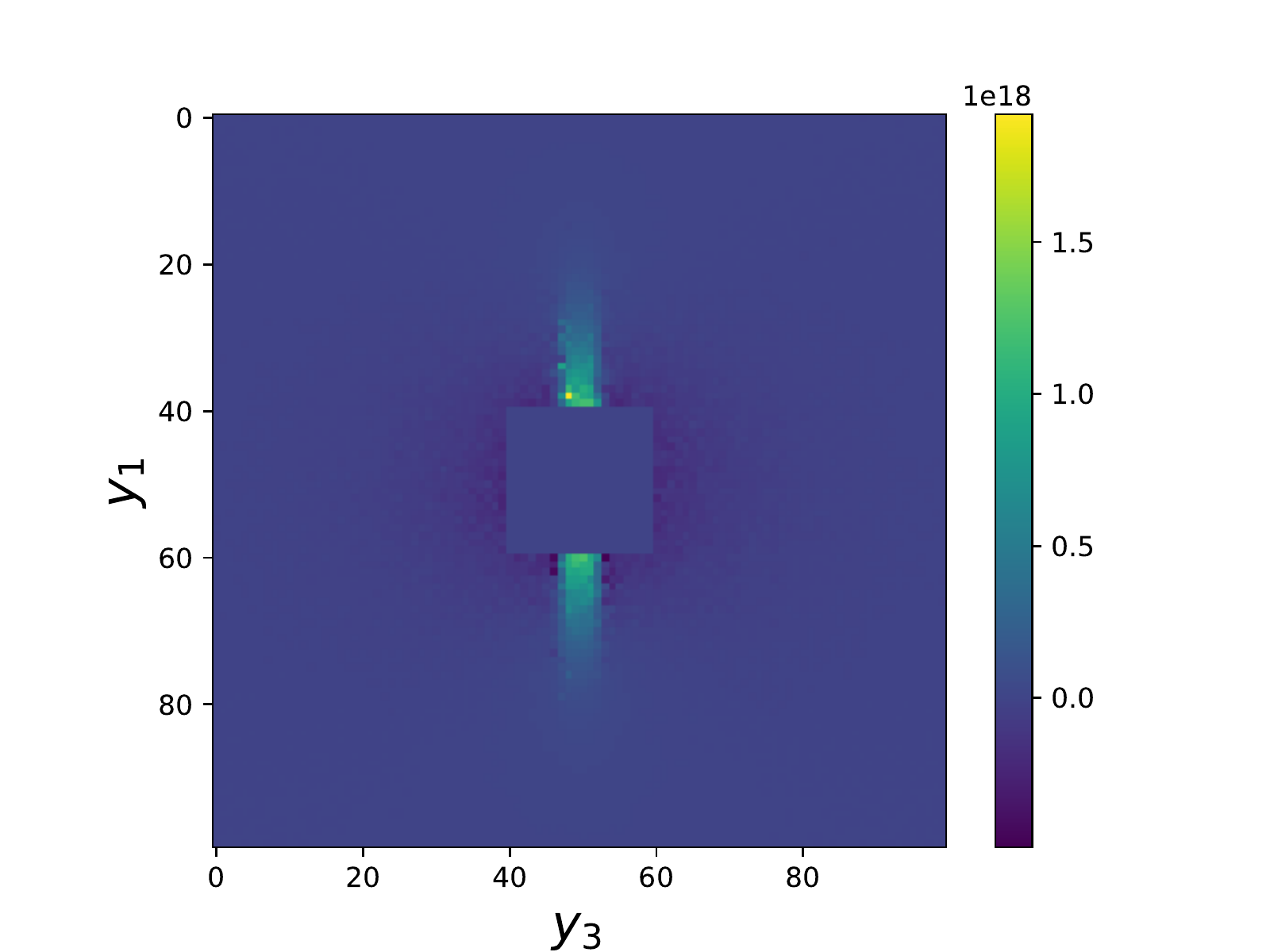}
\caption{This figure can be compared with Fig.\ \ref{disco0-10-y3-zero}, but with source-sink having diameter $D=10\cdot 10^{-7}$ cm.
} 
\label{disco10-50-y1-zero}
  \end{center}
\end{figure}

While we sample the integrand in order to evaluate the total integral, we also make a map of the function resulting from the partial integration in $\mathbf{z}$. For this purpose we divide the 3D cube of integration in $\mathbf{y}$ (with variable side $2R_y$) into 100 cells along each direction. The contribution to the magnetic field from each cell can be interpreted as being generated by a secondary current, or ``cloud'' current, proportional to the gradient of the field $S$. As seen from Tables \ref{table1} and \ref{table2} and from Figures \ref{puntif0-10}-\ref{disco10-50-y1-zero}, the regions with the largest secondary current density are those between the current source and sink and close to them, but the large cloud that extends around them and whose density decreases slowly with distance also gives a relevant contribution to the total integral; in the end, the latter contribution exactly cancels that of the localized sources. In order to evaluate the far contributions we sample the integrand over cubes of $\mathbf{y}$ of increasing size, excluding each time a cubic core equal to the previous cube. This is necessary because the number of sampling points cannot be increased beyond a practical limit of the order or $10^{10}$, and if we would sample the cores together with the periphery, the cores would produce too much noise. All the physical parameters have been chosen of the same magnitude order as in \cite{modanese2019design}.

\begin{table}
\begin{center}
\begin{tabular}{|c|c|} 
\toprule
range $R_y$ (units $10^{-7}$) & contribution to $B^s/B^0$ \nonumber \\
\hline
$[0-10]$ & $0.338$ \nonumber \\
$[10-50]$ & $-0.005$ \nonumber \\
$[50-100]$ & $-0.001$ \nonumber \\
$[100-200]$ & $-0.004$ \nonumber \\
$[200-400]$ & $-0.001$ \nonumber \\
$[400-800]$ & $-0.036\pm 0.003$ \nonumber \\
$[800-1100]$ & $-0.09\pm 0.01$ \nonumber \\
$[1100-1600]$ & $-0.133$ \nonumber \\
$[1600-3200]$ & $-0.046$ \nonumber \\
$[3200-6400]$ & $-0.006$ \nonumber \\
Total & $0.016\pm 0.014$ \nonumber \\
\hline
\end{tabular}	
\caption{Contributions of the secondary current $\nabla S$, in various regions of the $y$-space, to the ratio $B^s/B^0$. $B^s$ is the anomalous field generated by the secondary current. $B^0$ is the Biot-Savart field that the same total current would generate if flowing as a primary local current through the gap of length $2a$. The field is computed at a distance $r=10^{-4}$ cm from the origin, on the $x_1$ axis. The shape of the current sink and source is spherical ($\varepsilon=D=0.5\cdot 10^{-7}$ cm). The integration range in $z_1$ and $z_2$ is $R=2\cdot 10^{-7}$ cm, in $z_3$ is $R_3=5\cdot 10^{-7}$ cm. In the $y$ range $[800-1100]\cdot 10^{-7}$ cm, which contains the point where $B^s$ is computed, the integration is further split into two parts in order to reduce the noise due to the factor $1/|\mathbf{x}-\mathbf{y}|$. As seen from the last row of the table, the total anomalous field is zero within errors (``missing field'' effect). 
}		
\label{table1}
\end{center}
\end{table}

\begin{table}
\begin{center}
\begin{tabular}{|c|c|} 
\toprule
range $R_y$ (units $10^{-7}$) & contribution to $B^s/B^0$ \nonumber \\
\hline
$[0-10]$ & $0.324$ \nonumber \\
$[10-50]$ & $0.009$ \nonumber \\
$[50-100]$ & $-0.000$ \nonumber \\
$[100-200]$ & $-0.003$ \nonumber \\
$[200-400]$ & $-0.014$ \nonumber \\
$[400-800]$ & $-0.042\pm 0.003$ \nonumber \\
$[800-1100]$ & $-0.088\pm 0.005$ \nonumber \\
$[1100-1600]$ & $-0.133$ \nonumber \\
$[1600-3200]$ & $-0.046$ \nonumber \\
$[3200-6400]$ & $X$ \nonumber \\
Total & $0.007+X\pm 0.010$ \nonumber \\
\hline
\end{tabular}	
\caption{Same as in Tab.\ \ref{table1}, but for a sink and source with the shape of a disk/ellipsoid ($\varepsilon=0.5\cdot 10^{-7}$ cm, $D=2.5\cdot 10^{-7}$ cm). The integration range in $z_1$ and $z_2$ changes accordingly: here $R=10\cdot 10^{-7}$ cm. The total anomalous field is again zero within errors; the single contributions differ significantly from the case of spherical sink and source only at small $R_y$ range. The data for a wider disk ($D=10\cdot 10^{-7}$ cm, not shown here) reveal a similar behavior: in the first two $y$ ranges we have respectively contributions 0.185 and 0.143, with no substantial differences in the other ranges.
}		
\label{table2}
\end{center}
\end{table}

\section{Conclusion}

In summary, we have shown on general grounds that for any charge density $\rho$ and current density $\mathbf{j}$ given by a microscopic model, the generated e.m.\ field is given by eqs.\ (\ref{wave-E}) and (\ref{wave-B}), no matter if the continuity equation is satisfied or not. 

In the case of oscillating charges and currents, the radiated longitudinal electric field is given by eq.\ (\ref{E-Fourier}) or (\ref{E-time}). For example, if a fraction $\eta$ of a charge $q$ oscillating along the $z$-axis over a distance $2a$ lacks a corresponding current, the predicted maximum longitudinal field (along $z$) is $E_{L,max}=2\eta\omega^2 qa/(c^2r)$. It is tempting to speculate that this kind of oscillations could happen in graphene-based nanojunctions or carbon nano-wires \cite{wang2013time,yu2014current,walz2015local,pohl2019imaging,garner2020three}. 

The equation for the magnetic field (\ref{wave-B}) indicates that
it is possible in principle to observe experimentally if a fraction $\eta$ of a stationary current flows in a conductor with microscopic spatial ``interruptions'' affecting a volume that is a fraction $\chi$ of the total volume: in that case, the average Biot-Savart magnetic field generated will be $(1-\eta\chi)B_0$, where $B_0$ is the field generated by the same total current flowing in a conductor where $\nabla \cdot \mathbf{j}=0$ everywhere. The microscopic models cited in this work allow in principle to estimate the fractions $\eta$ and $\chi$. On the experimental side, a possible strategy to prove some anomalies (though not to disprove them) consists of looking for changes in the ratio $B/i$ in dependence on $i$. We have briefly discussed the possible uncertainties associated with this measurement.

Finally we note that the control of magnetic field patterns in molecular devices may be interesting for applications to high-density memory storage consisting of single molecular magnets (see \cite{nozaki2017current} and refs.). It has been further suggested that NMR-type experiments can be performed in order to observe the spatial fluctuations of magnetic fields generated by DC current flow \cite{walz2014current}.

\appendix

\section{Equations in SI units}
In SI units the equations for the potentials and the auxiliary field $S$ read
\begin{eqnarray*}
\frac{1}{c^{2}}\frac{\partial ^{2}\phi }{\partial t^{2}}-\nabla ^{2}\phi &=&%
\frac{\rho }{\varepsilon _{0}}-\frac{\partial S}{\partial t}, \\
\frac{1}{c^{2}}\frac{\partial ^{2}\mathbf{A}}{\partial t^{2}}-\nabla ^{2}%
\mathbf{A} &=&\mu _{0}\mathbf{j}+\nabla S, \\
\frac{1}{c^{2}}\frac{\partial ^{2}S}{\partial t^{2}}-\nabla ^{2}S &=&\mu _{0}%
\left[ \frac{\partial \rho }{\partial t}+\nabla \cdot \mathbf{j}\right]
\equiv \mu _{0}I,
\end{eqnarray*}
with $\varepsilon_{0}$ and $\mu_{0}$ the permittivity and permeability of free space, respectively.

The EM fields are expressed as
\begin{eqnarray*}
\mathbf{E} &=&-\nabla \phi -\frac{\partial \mathbf{A}}{\partial t}, \\
\mathbf{B} &=&\nabla \times \mathbf{A}.
\end{eqnarray*}

For a localized extra-source $I$, the solution for $S$ is%
\begin{equation*}
S\left( \mathbf{x},t\right) =\frac{\mu _{0}}{4\pi }\int \frac{I\left( 
\mathbf{x}^{\prime },t^{\prime }\right) }{\left\vert \mathbf{x}-\mathbf{x}%
^{\prime }\right\vert }d^{3}x^{\prime },
\end{equation*}%
with $t^{\prime }=t-\left\vert \mathbf{x}-\mathbf{x}^{\prime }\right\vert /c$

The wave equations for the electric and
magnetic fields are:
\begin{eqnarray*}
\frac{1}{c^{2}}\frac{\partial ^{2}\mathbf{E}}{\partial t^{2}}-\nabla ^{2}%
\mathbf{E} &=&-\mu _{0}\frac{\partial \mathbf{j}}{\partial t}-\frac{\nabla
\rho }{\varepsilon _{0}}=-\mu _{0}\left( \frac{\partial \mathbf{j}}{\partial
t}+c^{2}\nabla \rho \right) , \\
\frac{1}{c^{2}}\frac{\partial ^{2}\mathbf{B}}{\partial t^{2}}-\nabla ^{2}%
\mathbf{B} &=&\mu _{0}\nabla \times \mathbf{j}.
\end{eqnarray*}

For localized sources the corresponding solutions are%
\begin{eqnarray*}
\mathbf{E}\left( \mathbf{x},t\right) &=&-\frac{\mu _{0}}{4\pi }\int \left( 
\frac{\partial \mathbf{j}}{\partial t^{\prime }}+c^{2}\nabla ^{\prime }\rho
\right) \frac{d^{3}x^{\prime }}{\left\vert \mathbf{x}-\mathbf{x}^{\prime
}\right\vert }, \\
\mathbf{B}\left( \mathbf{x},t\right) &=&\frac{\mu _{0}}{4\pi }\int \nabla
^{\prime }\times \mathbf{j}\frac{d^{3}x^{\prime }}{\left\vert \mathbf{x}-%
\mathbf{x}^{\prime }\right\vert }.
\end{eqnarray*}

\section{Alternative proof of the independence of the EM fields on the
scalar source.}
\label{alternative}

The contribution to the potentials given by the scalar source is%
\begin{eqnarray*}
\phi \left( \mathbf{x},t\right) &=&-\frac{1}{4\pi c}\int \frac{\partial
S\left( \mathbf{x}^{\prime },t^{\prime }\right) }{\partial t^{\prime }}\frac{%
1}{\left\vert \mathbf{x}-\mathbf{x}^{\prime }\right\vert }d^{3}x^{\prime },
\\
\mathbf{A}\left( \mathbf{x},t\right) &=&\frac{1}{4\pi }\int \nabla ^{\prime
}S\left( \mathbf{x}^{\prime },t^{\prime }\right) \frac{\mathbf{1}}{%
\left\vert \mathbf{x}-\mathbf{x}^{\prime }\right\vert }d^{3}x^{\prime },
\end{eqnarray*}%
where it is assumed that at infinity no scalar field exists (no incoming
scalar field from large distances in the remote past, and local additional
sources acting during finite time intervals).

Considering that%
\begin{eqnarray*}
\nabla \phi \left( \mathbf{x},t\right) &=&-\frac{1}{4\pi c}\int \frac{%
\partial S\left( \mathbf{x}^{\prime },t^{\prime }\right) }{\partial
t^{\prime }}\nabla \left( \frac{1}{\left\vert \mathbf{x}-\mathbf{x}^{\prime
}\right\vert }\right) d^{3}x^{\prime } \\
&=&\frac{1}{4\pi c}\int \frac{\partial S\left( \mathbf{x}^{\prime
},t^{\prime }\right) }{\partial t^{\prime }}\nabla ^{\prime }\left( \frac{1}{%
\left\vert \mathbf{x}-\mathbf{x}^{\prime }\right\vert }\right)
d^{3}x^{\prime } \\
&=&\frac{1}{4\pi c}\int \nabla ^{\prime }\left( \frac{\partial S\left( 
\mathbf{x}^{\prime },t^{\prime }\right) }{\partial t^{\prime }}\frac{1}{%
\left\vert \mathbf{x}-\mathbf{x}^{\prime }\right\vert }\right)
d^{3}x^{\prime } \\
&&-\frac{1}{4\pi c}\int \frac{1}{\left\vert \mathbf{x}-\mathbf{x}^{\prime
}\right\vert }\nabla ^{\prime }\left( \frac{\partial S\left( \mathbf{x}%
^{\prime },t^{\prime }\right) }{\partial t^{\prime }}\right) d^{3}x^{\prime }
\\
&=&\frac{1}{4\pi c}\oint \frac{\partial S\left( \mathbf{x}^{\prime
},t^{\prime }\right) }{\partial t^{\prime }}\frac{1}{\left\vert \mathbf{x}-%
\mathbf{x}^{\prime }\right\vert }d\mathbf{S}^{\prime } \\
&&-\frac{1}{4\pi c}\int \frac{1}{\left\vert \mathbf{x}-\mathbf{x}^{\prime
}\right\vert }\nabla ^{\prime }\frac{\partial S\left( \mathbf{x}^{\prime
},t^{\prime }\right) }{\partial t^{\prime }}d^{3}x^{\prime },
\end{eqnarray*}%
that%
\begin{eqnarray*}
\frac{\partial \mathbf{A}\left( \mathbf{x},t\right) }{\partial t} &=&\frac{1%
}{4\pi }\int \frac{\partial }{\partial t}\nabla ^{\prime }S\left( \mathbf{x}%
^{\prime },t^{\prime }\right) \frac{\mathbf{1}}{\left\vert \mathbf{x}-%
\mathbf{x}^{\prime }\right\vert }d^{3}x^{\prime } \\
&=&\frac{1}{4\pi }\int \frac{\mathbf{1}}{\left\vert \mathbf{x}-\mathbf{x}%
^{\prime }\right\vert }\frac{\partial }{\partial t^{\prime }}\nabla ^{\prime
}S\left( \mathbf{x}^{\prime },t^{\prime }\right) d^{3}x^{\prime },
\end{eqnarray*}%
and that%
\begin{eqnarray*}
\nabla \times \mathbf{A}\left( \mathbf{x},t\right) &=&\frac{1}{4\pi }\int
\nabla ^{\prime }S\left( \mathbf{x}^{\prime },t^{\prime }\right) \nabla
\times \left( \frac{\mathbf{1}}{\left\vert \mathbf{x}-\mathbf{x}^{\prime
}\right\vert }\right) d^{3}x^{\prime } \\
&=&-\frac{1}{4\pi }\int \nabla ^{\prime }S\left( \mathbf{x}^{\prime
},t^{\prime }\right) \nabla ^{\prime }\times \left( \frac{\mathbf{1}}{%
\left\vert \mathbf{x}-\mathbf{x}^{\prime }\right\vert }\right)
d^{3}x^{\prime } \\
&=&\frac{1}{4\pi }\int \nabla ^{\prime }\times \left[ \nabla ^{\prime
}S\left( \mathbf{x}^{\prime },t^{\prime }\right) \frac{\mathbf{1}}{%
\left\vert \mathbf{x}-\mathbf{x}^{\prime }\right\vert }\right]
d^{3}x^{\prime } \\
&&-\frac{1}{4\pi }\int \frac{\mathbf{1}}{\left\vert \mathbf{x}-\mathbf{x}%
^{\prime }\right\vert }\nabla ^{\prime }\times \left[ \nabla ^{\prime
}S\left( \mathbf{x}^{\prime },t^{\prime }\right) \right] d^{3}x^{\prime }, \\
&=&-\frac{1}{4\pi }\oint \frac{\mathbf{1}}{\left\vert \mathbf{x}-\mathbf{x}%
^{\prime }\right\vert }\nabla ^{\prime }S\left( \mathbf{x}^{\prime
},t^{\prime }\right) \times d\mathbf{S}^{\prime },
\end{eqnarray*}%
we have for the EM fields%
\begin{eqnarray*}
\mathbf{E}\left( \mathbf{x},t\right) &=&-\nabla \phi -\frac{1}{c}\frac{%
\partial \mathbf{A}}{\partial t}=-\frac{1}{4\pi c}\oint \frac{1}{\left\vert 
\mathbf{x}-\mathbf{x}^{\prime }\right\vert }\frac{\partial S\left( \mathbf{x}
^{\prime },t^{\prime }\right) }{\partial t^{\prime }}d\mathbf{S}^{\prime },
\\
\mathbf{B}\left( \mathbf{x},t\right) &=&\nabla \times \mathbf{A}=-\frac{1}{%
4\pi }\oint \frac{\mathbf{1}}{\left\vert \mathbf{x}-\mathbf{x}^{\prime
}\right\vert }\nabla ^{\prime }S\left( \mathbf{x}^{\prime },t^{\prime
}\right) \times d\mathbf{S}^{\prime }.
\end{eqnarray*}

Since the scalar field is assumed to originate from a localized source that
was not acting in the infinite past, the surface integrals at infinity are
zero for finite $\left( \mathbf{x},t\right) $. This is so because the
contributions from $\left\vert \mathbf{x}^{\prime }\right\vert \rightarrow
\infty $ to the fields at finite values of $\left( \mathbf{x},t\right) $,
come from the scalar at $t^{\prime }\rightarrow -\infty $, which is zero by hypothesis.

\bibliographystyle{unsrt}
\bibliography{mme}

\end{document}